\documentclass[twocolumn,pra,aps]{revtex4}
\usepackage{graphicx}
\usepackage{amsmath}

\begin{document}
\title{Quantum communications with time-bin entangled photons:\\
long distance quantum teleportation and quantum repeaters}
\author{N. Gisin$^\dag$, I. Marcikic$^\dag$, H. de Riedmatten$^\dag$, W.
Tittel$^{\dag,\ddag}$ and H. Zbinden$^\dag$}

\affiliation{$^\dag$Group of Applied Physics-Optique, University of
Geneva, CH-1211, Geneva 4,
Switzerland\\
$^\ddag$QUANTOP - Danish National Research Foundation Center for
Quantum Optics, Institute for Physics and Astronomy, University of
Aarhus, Denmark}

\begin{abstract}
Using 2 km of standard telecom optical fibres, we teleport qubits
carried by photons of 1.3 $\mu$m wavelength to qubits in another
lab carried by a photons of 1.5 $\mu$m wavelength. The photons to
be teleported and the necessary entangled photon pairs are created
in two different non-linear crystals. The measured mean fidelity
is of 81.2\%. We discuss how this could be used as quantum
repeaters without quantum memories.
\end{abstract}

\maketitle
\section{Introduction}
Long distance quantum communication in optical fibers should
exploit the standard and widely installed telecom fibers. This
standard imposes the wavelength (either around 1300 nm, or around
1550 nm). It does also strongly suggest that polarization encoding
is not the optimal choice. Indeed, polarization effects in telecom
fibers fluctuate on a time scale from ms to tens of minutes for
aerial and underground cables, respectively. These unavoidable
fluctuations require active feedback and/or compensation schemes.
An alternative consists in encoding the qubits in time-bins.
Time-bin qubits have already been used for quantum
cryptography\cite{TittelQC}, for the production of non-maximally
entangled qubits\cite{ThewRobust} and to encode quantum states in
higher dimensional Hilbert spaces\cite{HighDim}. In this
contribution we present the results of an experimental
demonstration of long distance quantum teleportation, using
time-bin qubits. In section \ref{sec:Qrepeater} we present a
potential application of quantum teleportation as "quantum
repeaters without quantum memory".\\
\begin{figure}[h]
\includegraphics[width=8.43cm]{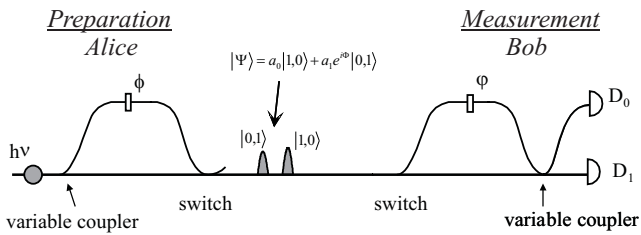}
\caption{Preparation and measurement of time bin qubits}
\label{timebin}
\end{figure}

Figure \ref{timebin} presents a general preparation and
measurement scheme for time-bin qubits\cite{TittelWeihs}. In
principle, any state can be deterministically prepared and ideal
measurements in any basis performed. In practice, however, the
switches are replaced by passive couplers, hence induce 50\% loss.
Also, in our experiment we replaced the variable coupler either by
a passive coupler (corresponding to a coupling ratio of 50\%), or
by fibers of appropriate lengths (corresponding to 0\% and 100\%
coupling ratios)\ref{comment1}.

\section{Quantum teleportation}\label{sec:Qtelep}

\begin{figure}[h]
\includegraphics[width=8.43cm]{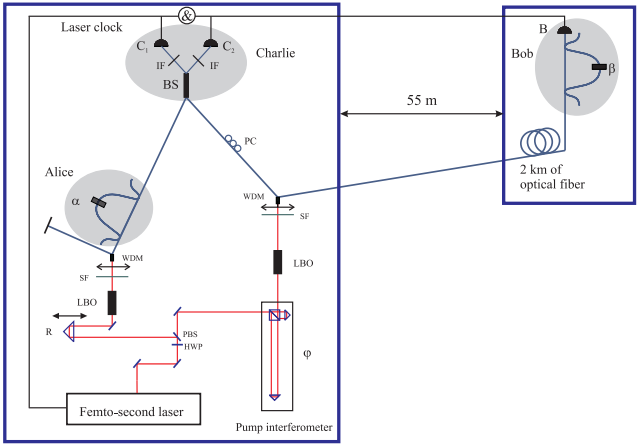}
\caption{Experimental setup} \label{setup}
\end{figure}

Our experiment is schematically presented on Fig. \ref{setup}. For
a general introduction to quantum teleportation and previous
experiments, we refer
to\cite{BennetTelep,ViennaTelep,RomeTelep,ShiTelep,KimbleTelep}.

A mode locked Ti:Sapphire laser produces 150 fs pulses at
$\lambda=710$ nm. The laser beam is split into two parts. The
transmitted beam is used to create entangled time-bin qubits (EPR
source) by passing the beam first through an unbalanced bulk
Michelson interferometer with optical path-length difference
$\Delta\tau=1.2$ ns and then through a type I non-linear crystal
(LBO) where entangled non-degenerate collinear time-bin qubits at
telecom wavelengths (1310 and 1550 nm) are created. The pump light
is removed with a silicon filter (SF) and the twin photons are
collimated into an optical fibre and separated by a
wavelength-division-multiplexer (WDM). The 1310 nm photon is then
sent to Charlie and its twin 1550nm photon to Bob. The entangled
state is described by
$|1,0\rangle_{Charlie}\otimes|1,0\rangle_{Bob}+|0,1\rangle_{Charlie}\otimes|0,1\rangle_{Bob}$,
where $|1,0\rangle$ and $|0,1\rangle$ represents the first and
second time-bin, respectively.

The reflected beam is used to create the qubits to be teleported.
Similar to the creation of the entangled pairs, the beam is
focussed into a LBO crystal creating pairs of photons at 1310 nm
and 1550 nm wavelengths. After blocking the pump light and
coupling the photon pairs into an optical fibre, the 1550 nm
photon is removed using a WDM. Alice passes the 1310 nm photon
through a qubit preparation device, thereby creating a state
$|\psi_{Alice}\rangle=a_0|1,0\rangle+a_1 e^{i\alpha}|0,1\rangle$,
where $a_0=0,1$ or $1/\sqrt{2}$ and $a_{0}^{2}+a_{1}^{2}=1$.
Alice's qubit is finally sent to Charlie.
Charlie performs the joint Bell-state measurement between the
qubit sent by Alice and his part of the pair, with the 50/50 fibre
coupler BS. We select only the projection onto the singlet
entangled state
$|\psi^{(-)}\rangle=|1,0\rangle_{Alice}\otimes|0,1\rangle_{Charlie}-|0,1\rangle_{Alice}\otimes|1,0\rangle_{Charlie}$.
This takes place when the two photons trigger the detectors
labelled C1 and C2 on Fig. 2 at times that differ precisely by the
time difference $\Delta\tau$ between two time-bins. Detectors C1
and C2 are a $LN_{2}$ cooled passively quenched Ge avalanche
photodiode (APD) and a Peltier cooled InGaAs APD
respectively\cite{idQ}. Bob is situated in another lab, 55 m away
from Charlie. To simulate a longer distance we added 2 km of
optical fibre before the teleported photon reaches Bob's analyser.

The quality of the teleportation is usually reported in terms of
fidelity $\bar F$, i.e. the probability that Bob's qubit
successfully passes an analyser testing that it is indeed in the
state $\psi_{Alice}$  prepared by Alice, averaged over all
possible $\psi_{Alice}$. The linearity of quantum mechanics
implies:
\begin {equation}
\bar{F} = \int \langle\psi_{Alice}|\rho_{Bob}|\psi_{Alice}\rangle
d\psi_{Alice}=\frac{2}{3}F_{equator}+\frac{1}{3}F_{pole}
\end{equation}
where $F_{equator}$ and $F_{pole}$, are the fidelities averaged
only over the states corresponding to the equator and the poles of
the Poincar\'e sphere, respectively.
\begin{figure}[h]
\includegraphics[width=8.43cm]{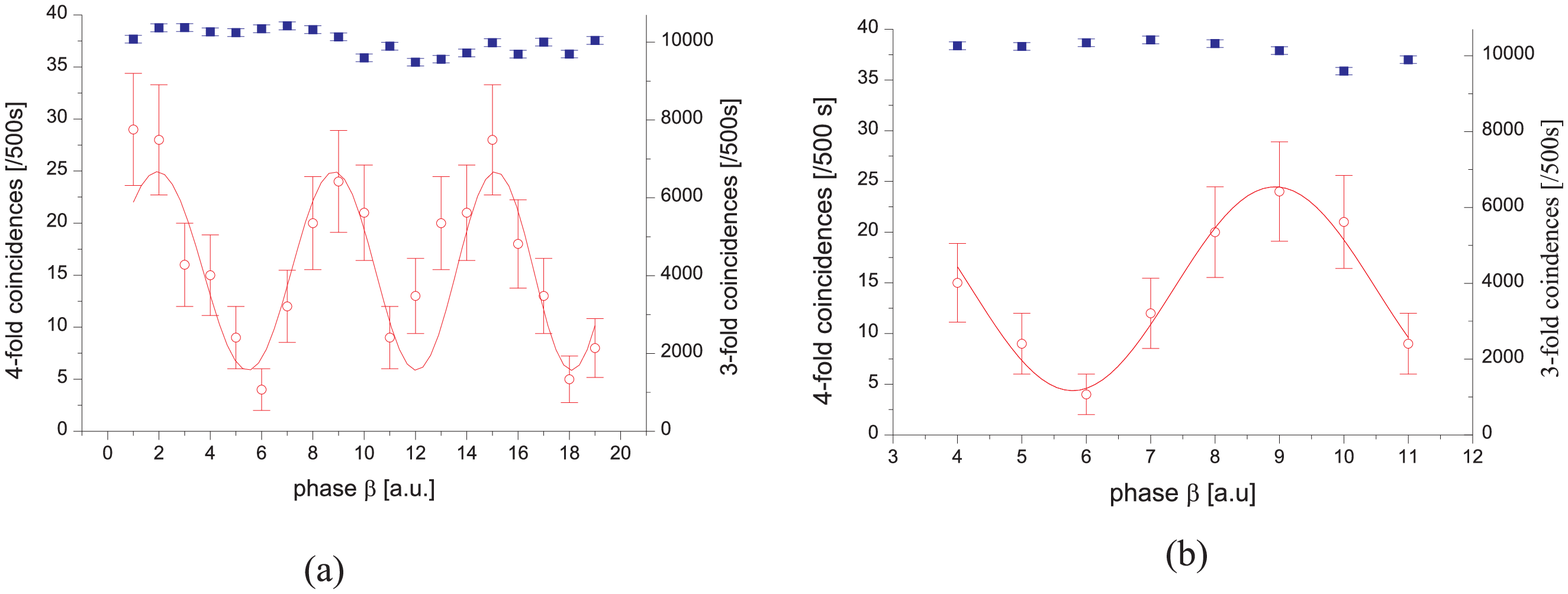}
\caption{Teleportation of equatorial states. Open circles are
4-fold coincidences (3 photons + laser clock), i.e detection of
Bob's photon conditioned on a successful Bell measurement. Black
squares are 3-fold coincidences (Photons C1+B+ laser clock). They
should remain constant, since they contains no information about
the Bell measurement} \label{results}
\end{figure}

In our experiment we measured $F_{pole}$ directly, by preparing
and measuring each of the two corresponding states:
$F_{pole}=(82.5 \pm 3)\%$. In order to measure $F_{equator}$, we
prepared various states, using a 50/50\% coupler with various
phases $\alpha$, and for each scanned the analyzing
interferometer's phase $\beta$. From the raw visibility $V_{raw}$
one obtains the fidelity $F=\frac{1+V_{raw}}{2}$. Typically, we
obtain $F_{equator}= (80.5 \pm 2.5)\%$(Fig. 3a), with best results
achieving values up to $(85 \pm 2.5)\%$ (Fig. 3b). Accordingly,
the measured fidelity, averaged over all possible states
$\psi_{Alice}$ equals $\bar F=(81.2 \pm 2.5)\%$.

\section{Quantum repeaters without quantum memories}\label{sec:Qrepeater}
Generally, quantum channels are lossy, hence the probability that
a single photon sent by Alice reaches Bob decreases with distance.
Because of the no-cloning theorem, there is no way to merely
amplify the signal. With noise-free detectors, the attenuation
would only affect the bit rate, and protocols like quantum
cryptography would provide secure keys for arbitrary distances.
However, realistic detectors have a finite probability of dark
counts (around $10^{-4}$ per ns for InGaAs APDs\cite{idQ}). This
noise is independent of the distance, hence there is a limiting
distance after which the signal becomes smaller than the noise. In
a series of papers, H. Briegel and co-workers\cite{Briegel} have
shown that the appropriate use of entanglement purification and of
quantum memories allows one, in principle, to realize quantum
repeaters which would extend the range of quantum cryptography to
unlimited distances, with only a polynomial increase of repeater
stations. This is a promising line of research for experimental
physics, but it is only fair to say that it will remain a research
topic for the years to come.
\begin{figure}[h]
\includegraphics[width=8.43cm]{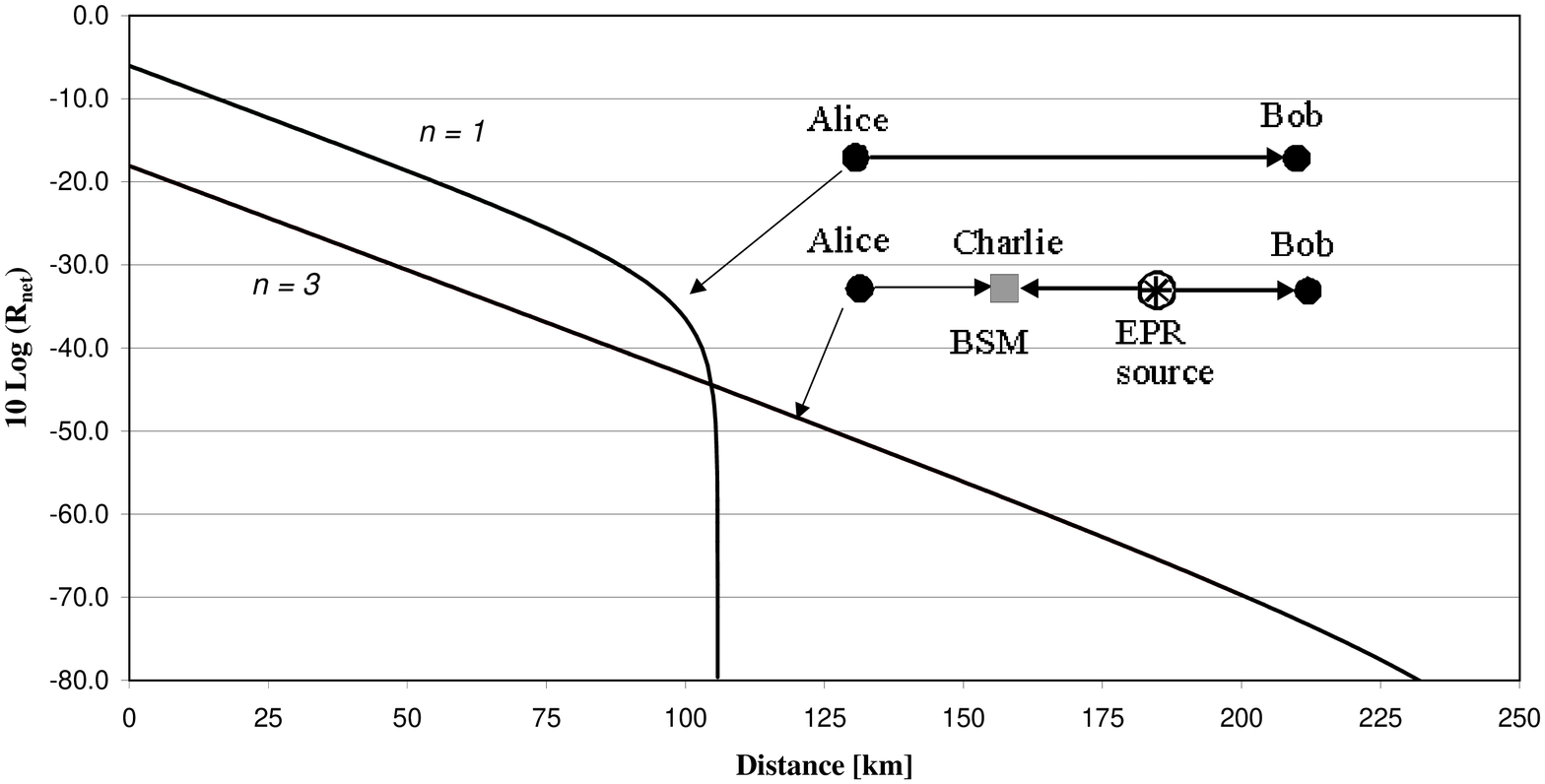}
\caption{Net count rate as a function of the distance when Alice
is directly connected to Bob (n=1) and in the case of
teleportation (n=3)} \label{Rnet}
\end{figure}

However, without quantum memories and using only near future
technologies (like the ones presented in this contribution), there
are ways to extends the distances over which secret quantum keys
can be distributed, although not to unlimited
distances\cite{rmp,YamamotoQR,FransonQR}. Assume that Alice and
Bob are connected by a quantum communication channel with
transmission coefficient $t$. Denote $\eta$ the photon counting
detector's efficiency and $D$ the dark count probability per
transmission (i.e. per qubit). Furthermore, denote $C$ and $Q$ the
rates of correctly and incorrectly detected photons: In good
approximation\footnote{We checked that a complete analysis doesn't
change qualitatively the results}, we have $C=t\eta$,
$Q=(1-t\eta)D$. Accordingly, the total rate of transmission
$R_{raw}$ and the QBER (Quantum Bit error Rate) read:
\begin{equation}
R_{raw}=C+Q \hspace{2cm} QBER=\frac{Q}{C+Q}
\end{equation}
 The net
rate $R_{net}$, after error correction and privacy amplification,
is proportional to the raw rate $R_{raw}$ with a coefficient that
depends on the $QBER$ \cite{rmp}. This coefficient is the
difference between the Alice-Bob $I_{AB}$ and the Alice-Eve
$I_{AE}$ mutual Shannon informations (or the Bob-Eve one, if
smaller). The exact relation between $I_{AE}$ and the QBER depends
on the exact eavesdropping model. Our results are essentially
independent of the details, though numerical values may slightly
differ. For the optimal individual attack, $I_{AB}-I_{AE}$ is
almost a linear function of the QBER, hence we use the
approximation:
\begin{equation}
R_{net}\approx R_{raw}\cdot(1-\frac{QBER}{15\%})=C-\frac{85}{15}Q
\end{equation}
 With realistic numbers, $\eta=0.25$, $D=10^{-4}$, and an
attenuation coefficient $\alpha=0.25$ dB/km (i.e. $t=10^{-\alpha
L/10}$, $L=$ fiber length), this sets an absolute limit for
quantum key distribution around 105 km (even with perfect single
photon sources), see Fig. 4. Consider now the case where the
channel is divided into 3 sections, each of length $L/3$, i.e.
with transmission coefficient $t^{1/3}$ (Fig. 4), exactly as in a
quantum teleportation experiments. In such a configuration, Bob
correctly detects the qubit whenever all 3 photons make it to
their detectors, all 3 detectors click and the Bell measurement is
successful, i.e. $C^{(3)}=\frac{1}{2}(t^{1/3}\eta)^3=
\frac{1}{2}t\eta^3$. The rate of incorrect detections contains
three parts\footnote{We assumed that the detector noise completely
dominates the error rate. This is a reasonable assumption when
Alice is directly connected to Bob (i.e. when $n=1$),
see\cite{stucki02}. But, clearly, when the systems gets more
complex, optical alignment is less optimal. How this optical
contribution to the error rate scales with $n$ depends on the
detailed implementation. From our experiment we believe that the
"optical error rate" can be maintained quite low.}: (i) all 3
photons are lost and there are 3 dark counts, (ii) 2 photons are
lost, 1 detected and there are 2 dark counts, (iii) 1 photon is
lost, 2 detected and there is one dark count. Accordingly:
\begin{center}
\begin{eqnarray}
 \nonumber Q^{(3)}&=&(1-t^{1/3}\eta)^3\cdot
D^3+3t^{1/3}\eta(1-t^{1/3}\eta)^2\cdot D^2
\\\nonumber&&+ 3(t^{1/3}\eta)^2(1-t^{1/3}\eta)\cdot D\\
&=&\left(t^{1/3}\eta+(1-t^{1/3}\eta)D\right)^3 - t\eta^3
\label{R3}
\end{eqnarray}
\end{center}
 The corresponding net rate, with the
same parameters, is also displayed on Fig. 4. Clearly, the
division of the channel into 3 trunks and the use of teleportation
increases significantly the maximal distance over which quantum
key distribution is possible. The generalization to divisions into
n trunks is straightforward. For given parameters the limiting
distance increases with n, up to a maximum, and then decreases
again.
\section{Concluding remarks}\label{sec:concl}
Objects are constituted by energy and structure (or matter and
form according to Aristotle). In quantum teleportation one does
not teleport neither energy nor matter. However, the ultimate
structure of objects is indeed teleported from and place to
another, without the object ever being anywhere in between!
Admittedly, teleportation of systems much more complex than a
simple qubit is still elusive. But the teleportation of qubits
could already be useful for quantum cryptography, though much
effort still have to be devoted to improve the results of this
first long distance realization, especially the stability of the
experimental scheme. \\

Financial support by the Swiss OFES and NSF within the framework
of the European IST project Qucomm and the Swiss National Center
for Quantum Photonics is acknowledged. W.T. acknowledges funding
by the ESF Programme Quantum Information Theory and Quantum
Computation (QIT).


\begin{thebibliography}{99}
\bibitem{comment1} The replacement of the switches is necessary because real and
fast switches have losses even larger than 50$\%$. However, real
variable couplers could be used.
\end{thebibliography}
\end{document}